\documentclass[preprint,prb,showpacs,superscriptaddress,amsmath,citeautoscript]{revtex4}
\usepackage{graphicx}
\usepackage{bm}
\usepackage{amssymb}

\begin{document}

\title{Dynamic micro-Hall detection of superparamagnetic beads in a microfluidic channel }

\author{K.\ Aledealat}
\email{ka04c@fsu.edu}
\affiliation{Department of Physics and MARTECH, Florida State University, Tallahassee, FL 32306\\}
\author{G.\ Mihajlovi\'{c}}
\affiliation{Materials Science Division, Argonne National Laboratory, Argonne, IL 60439\\}
\author{K.-S.\ Chen}
\affiliation{Department of Physics and MARTECH, Florida State University, Tallahassee, FL 32306\\}
\author{M.\ Field}
\affiliation{Teledyne Scientific Company LLC, Thousand Oaks, CA 90360\\}
\author{G.\ J.\ Sullivan}
\affiliation{Teledyne Scientific Company LLC, Thousand Oaks, CA 90360\\}
\author{P.\ Xiong}
\affiliation{Department of Physics and MARTECH, Florida State University, Tallahassee, FL 32306\\}
\affiliation{Integrative NanoScience Institute, Florida State University, Tallahassee, FL 32306\\}
\author{P.\ B.\ Chase}
\affiliation{Department of Biological Science, Florida State University, Tallahassee, FL 32306\\}
\affiliation{Integrative NanoScience Institute, Florida State University, Tallahassee, FL 32306\\}
\author{S.\ von Moln\'{a}r}
\affiliation{Department of Physics and MARTECH, Florida State University, Tallahassee, FL 32306\\}
\affiliation{Integrative NanoScience Institute, Florida State University, Tallahassee, FL 32306\\}
\date{\today}


\begin{abstract}
We report integration of an InAs quantum well micro-Hall sensor with microfluidics and
real-time detection of moving superparamagnetic beads for biological applications. The
detected positive and negative signals correspond to beads moving within and around the
Hall cross area respectively. Relative magnitudes and polarities of the signals measured
for a random distribution of immobilized beads over the sensor are in good agreement
with calculated values and explain consistently the dynamic signal shape. The fast sensor
response and its high sensitivity to off-cross area beads demonstrate its capability for
dynamic detection of biomolecules and long-range monitoring of non-specific binding
events.
\end{abstract}

\maketitle

A wide range of potential applications in biology and medicine rely on using superparamagnetic (SPM) bead labels for monitoring and manipulating biological species such as genetic testing \cite{Lagae2005}, detection of viruses \cite{Lee2008}, magnetic separation of biomolecules \cite{Pankhurst2003}, or drug delivery \cite{Pankhurst2003}. Significant progress has been made recently in detecting the presence of such beads in a dry state with various solid-state magnetic sensors \cite{Megens2007, Ejsing2004, Li2003, Li2006,Besse2002, Mihajlovic2005, Florescu2008, MihajlovicAPL2007}. On the other hand, microfluidic devices have been used to guide, manipulate, and sort magnetic beads \cite{Choi2002, Janssen2008, Spero2008, Weddemann2009} and to provide proper environment for biochemical interactions \cite{Weibel2006}. Hence, the integration of magnetic sensors with microfluidics is the key for achieving portable, low cost, fast-response electronic chips that can be utilized in biomedical applications. So far, the research efforts in this direction have exclusively employed sensors based on metallic multilayer magnetoresistance technologies \cite{Loureiro2009, Liu2009, Shen2005, Thilwind2008, Roh2009, Osterberg2009}. These sensors offer high sensitivity but often suffer from nonlinear responses, and more importantly, saturation at very low fields. On the other hand, micro-Hall devices have linear output and are not limited in the magnitude of the external magnetic fields that can be applied to magnetize SPM beads, which is advantageous for biodetection and necessary in magnetic separation. In particular, devices based on InAs quantum well (QW) semiconductor heterostructures have been recently demonstrated with a superb signal-to-noise (S/N) ratio in detection experiments performed with micrometer and nanometer sized beads \cite{Florescu2008, Manandhar2009, MihajlovicJAP2007}. Furthermore, since these devices are sensitive to perpendicular component of the stray magnetic field generated by the beads, they can map out the trajectory of the moving beads. Here, we demonstrate the first integration of InAs QW micro-Hall sensors with microfluidic channels and detection of micron-sized SPM beads flowing through the channel at an average speed of 3~$\mu$m/sec. We show that high sensitivity of our sensors even to off-cross area beads can be utilized to monitor non-specific biomolecular binding over relatively large spatial area, thus providing more reliable quantitative detection of the specific binding events.

The sensor used in this work was fabricated in the form of  an array of six 1 $\times$ 1 $\mu$m$^2$ Hall crosses using photolithography and wet chemical etching from molecular beam epitaxy grown heterostructure consisting of GaAs substrate/GaAs buffer (100 nm)/AlAs (10 nm)/AlSb (105.4 nm)/Al$_{0.8}$Ga$_{0.2}$Sb (2000 nm)/ Al$_{0.7}$Ga$_{0.3}$Sb (200 nm)/AlSb (8 nm)/InAs (QW 12.5 nm)/AlSb (3 nm)/Te delta doping (0 nm)/ AlSb (10 nm)/GaSb (0.6 nm)/ In$_{0.5}$Al$_{0.5}$As (5 nm). The device was covered by a 70~nm thick SiO$_2$ layer grown by RF magnetron sputtering to provide electrical insulation. The Hall coefficient of the sensor measured at room temperature was 188~$\Omega$/T. The mobility and the density of the electrons in the InAs QW were $\mu$ = 1.4 $\times$~10$^4$~cm$^2$/Vs and $n$ = 3.3 $\times$~10$^{12}$~cm$^{-2}$ respectively.

To fabricate a microfluidic channel a piece of glass cover slip was cut in a rectangular shape and glued onto Si/SiO$_2$ wafer to form the master. Posts were attached to the mold to define access holes. The master and the posts were then covered with Polydimethysiloxane (PDMS) and cured in the oven at 70 $^o$C for five hours. Finally, the PDMS replica was peeled off and sealed to the sensor chip. The height and the width of the channel were 200 $\mu$m and 1000 $\mu$m respectively. The sensor and the PDMS channel were sandwiched inside a scaffold to strengthen the seal and prevent leakage. An optical image of the micro-Hall sensor integrated with microfluidics is shown in Fig. 1.

\begin{figure}
\includegraphics[scale=0.75, bb=6 7 323 180]{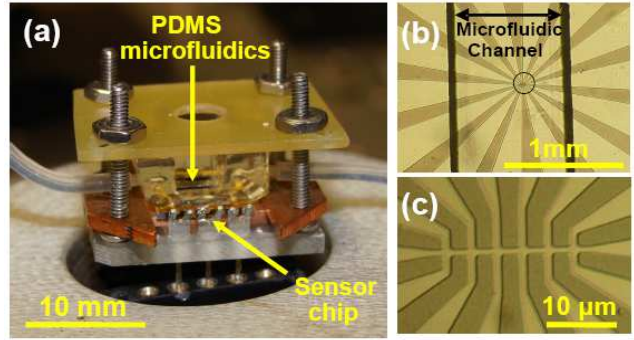}
\caption{(Color online) (a) Optical image of a micro-Hall sensor chip integrated with microfluidics. (b) A PDMS microfluidic channel aligned with the sensor chip. (c) Six 1 $\times$ 1 $\mu$m$^2$ Hall crosses located under the center of the microfluidic channel (area marked by the black circle in Fig. 1(b).}
\label{fig:image}
\end{figure}

Bead suspension was prepared by diluting 10 $\mu$L of streptavidin coated SPM beads (Bangs Laboratories, Inc.), 2.6 $\mu$m in diameter, into 0.7 mL of deionized water. Beads with the same functional group have been successfully used in detection of specific biomolecular interactions with micro-Hall magnetic sensors in a dry state \cite{Manandhar2009}. The fluid was pumped into the channel through microtubes using a syringe pump and was monitored via an optical microscope equipped with a video camera. Since the beads have higher density (1.2 g/cm$^3$) than water, the frictional force between the descended beads and the bottom of the channel is strong enough to slow the beads down to an average speed of 3 $\mu$m/s for a given flow rate of 1 $\mu$L/min.

In order to facilitate detection, an external $ac$ magnetic field of rms magnitude $B_0$ = 0.9~mT and frequency $f$ = 93~Hz was applied perpendicularly to the sensor plane to magnetize the beads. The sensor was biased with a $dc$ current $I$ = 50~$\mu$A and the Hall voltage $V_H$ was measured using a lock-in amplifier tuned to the frequency of the $ac$ magnetic field. As the beads pass by the cross they alter the total average magnetic field over its sensing area and hence $V_H$  associated with it. Due to dipolar character of the beads' stray magnetic field \cite{Besse2002, MihajlovicJAP2007}, the total field penetrating each cross will either increase or decrease depending on the relative position of the beads with respect to the cross. Fig. 2 shows a real-time $V_H$  signal recorded for one cross (the first one from the left in Fig. 2(b)). By matching the recorded movie of the flow of beads and the measured $V_H$, we find that the first dip in the exploded view of Fig. 2 (a) corresponds to a group of three beads approaching the sensing area of the Hall cross (Fig. 2 (b)). The successive peak appears when the first bead from the group passes right over the area (Fig. 2 (c)) while the second dip appears when the beads move away from the area (Fig. 2 (d)). By examining the rest of the data we find that the magnitudes of the recorded $V_H$ signals depend on the lateral distance between the beads and the Hall cross, the number of the passing beads, and their speed. In addition, we observed that each peak corresponds to beads passing over the sensing area and each dip corresponds to beads passing beside the area. In this measurement, the time constant of the lock-in amplifier was set to 300~ms at 12~dB/octave roll-off. The S/N ratio for the highest peak and the lowest dip recorded in the measurement were 22~dB and 24.7~dB respectively.

\begin{figure}
\includegraphics[scale=0.75, bb=15 3 320 307]{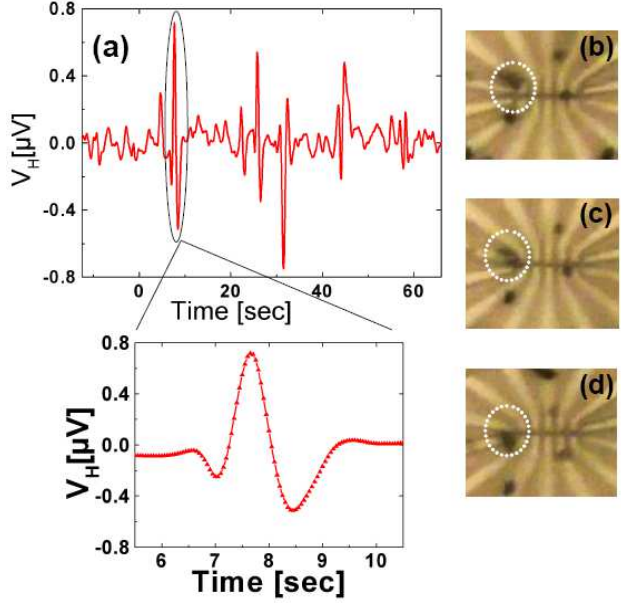}
\caption{(Color online) (a) $V_H$ as a function of time recorded during movement of SPM beads in a microfluidic channel. The exploded view shows the signal due to a line of three beads passing the measured cross (first one from the left on the snapshot series (b) to (d)). The two dips correspond to the line of beads approaching [image (b)] and leaving [image (d)] the cross while the middle peak appeared when  beads were partially covering the cross [image (c)].}
\label{fig:second}
\end{figure}

The conclusions drawn from the dynamic detection experiment were further confirmed in the experiment with immobilized beads. In this case, we prepared a drop of the same bead suspension on an identical Hall sensor and let it dry. Fig. 3 (a) shows an SEM image of the sensor. One bead partially covers one cross while the rest of the beads are surrounding the other crosses. A Hall detection measurement as described by Mihajlovi\'{c} et al. \cite{Mihajlovic2005, MihajlovicJAP2007} was then performed. Here, the main difference from the dynamic measurement is that an additional $dc$ magnetic field ($B$ = 70.6~mT) was applied in order to  reduce the induced $ac$ magnetization of the beads. This imitates the situation where the beads move away from the sensor. Fig. 3(b) shows the measured $V_H$ for four crosses shown in Fig. 3(a). The sign of $\Delta V_H$  signals agrees well with our interpretation of the dynamic measurement. Namely, for the empty crosses (1-3) $V_H$ went up, and for the occupied cross (4) down, when $B$  was applied. The magnitude of the $\Delta V_H$ signals can be also estimated by calculating the total average magnetic field penetrating sensing area of each cross due to nearby beads. Since the stray magnetic field from a SPM bead is dipolar \cite{Besse2002, MihajlovicJAP2007} and since the Hall sensor is only sensitive to the perpendicular component, we can explicitly write this average field as
\begin{equation}
  B_i=\frac{\mu_0 M V}{4 \pi}\sum_j \int\int\frac{2z^2-(x-x_j)^2-(y-y_j)^2}{(z^2+(x-x_j)^2+(y-y_j)^2)^{5/2}}dxdy.
  \label{eq:avfield}
\end{equation}
where indices $i$ and $j$ correspond to the Hall cross and the beads respectively, $M$ is the magnetization of the bead, $V$ its volume and the integration is performed over the sensing area of the Hall cross. We find that the calculated relative signals from crosses (2), (3) and (4) with respect to that of the cross (1) are [$B_2/B_1$, $B_3/B_1$, $B_4/B_1$] = [1.0, 0.9,-1.9]. The measured relative signals in the same order are [1.1, 0.8,-1.9]. The agreement is excellent, considering that the variations between the crosses and the beads were disregarded in calculations. Based on the measured noise level and according to this model, a single bead of the same size will be detectable within a radius of 4.5 $\mu$m from the center of the cross.

\begin{figure}
\includegraphics[scale=1, bb=6 5 214 267]{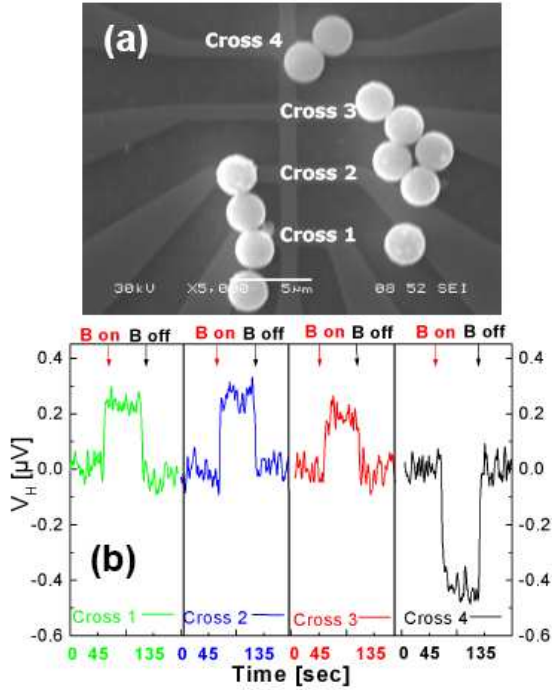}
\caption{(Color online) (a) An SEM image of the central region of an InAs micro-Hall sensor and immobilized SPM beads. Image shows one bead partially covering cross 4 and the rest of the beads are surrounding other crosses, (b) $V_H$ as a function of time for crosses 1 to 4. The increase of $V_H$ for crosses 1 to 3 and its drop for the cross 4 as $B$ was applied agree with the expected signals based on a dipolar stray field representation of SPM beads.}
\label{fig:third}
\end{figure}

The signal from off-cross area beads can be implemented in the detection of non-specific binding events in bio-assembly. For example, one  Hall cross can be functionalized to trap target molecules labeled by SPM beads, while other crosses in the vicinity can be passivated to serve as detectors of nonspecific bindings. A sensor array may be designed to have a biofunctionalized cross surrounded by several passivated crosses whose sensitivity range covers a continuous area that is significantly larger than that of the functionalized cross. Thus a signal from a functionalized cross can be more reliably interpreted as due to specific binding if signals from the passivated crosses are either negative or very low. This information would eliminate the necessity for post imaging (optical, fluorescent, or scanning electron microscopy) which is advantageous  for biomedical sensing that requires fast and reliable results.

In conclusion, we demonstrated real-time detection of micron-sized SPM beads using an InAs QW micro-Hall sensor integrated with a microfluidic channel. The recorded $V_H$ consists of peaks and dips that correspond to beads moving over the sensing area and around it respectively. The polarity and the magnitude of $\Delta V_H$ signals were successfully explained using a magnetic dipolar stray field representation of SPM beads. The high sensitivity of the Hall sensor to long range negative dipolar field generated by the beads can be utilized in monitoring the non-specific binding events and simultaneous analysis of the reliability of a biological assembly.

The authors would like to acknowledge the help of Kurt Koetz from Florida State University for his technical assistance in the initial stages of this work. The work was supported by NIH NIGMS GM079592.


\begin{thebibliography}{35}
\expandafter\ifx\csname natexlab\endcsname\relax\def\natexlab#1{#1}\fi
\expandafter\ifx\csname bibnamefont\endcsname\relax
  \def\bibnamefont#1{#1}\fi
\expandafter\ifx\csname bibfnamefont\endcsname\relax
  \def\bibfnamefont#1{#1}\fi
\expandafter\ifx\csname citenamefont\endcsname\relax
  \def\citenamefont#1{#1}\fi
\expandafter\ifx\csname url\endcsname\relax
  \def\url#1{\texttt{#1}}\fi
\expandafter\ifx\csname urlprefix\endcsname\relax\def\urlprefix{URL }\fi
\providecommand{\bibinfo}[2]{#2}
\providecommand{\eprint}[2][]{\url{#2}}


\bibitem[{\citenamefont{Lagae et~al.}(2005)\citenamefont{Lagae, L.; Wirix-Speetjens, R.; Liu, C.-X.; Laureyn, W.; Borghs, G.; Harvey, S.; Galvin, P.; Ferreira, H.A.; Graham, D.L.; Freitas, P.P.; Clarke, L.A.; Amaral, M.D.}}]{Lagae2005}
\bibinfo{author}{\bibfnamefont{L.}~\bibnamefont{Lagae}},
  \bibinfo{author}{\bibfnamefont{R.}~\bibnamefont{Wirix-Speetjens}},
  \bibinfo{author}{\bibfnamefont{C.~-X.} \bibnamefont{Liu}},
  \bibinfo{author}{\bibfnamefont{W.} \bibnamefont{Laureyn}},
  \bibinfo{author}{\bibfnamefont{G.} \bibnamefont{Borghs}},
  \bibinfo{author}{\bibfnamefont{S.} \bibnamefont{Harvey}},
  \bibinfo{author}{\bibfnamefont{P.} \bibnamefont{Galvin}},
  \bibinfo{author}{\bibfnamefont{H.~A.} \bibnamefont{Ferreira}},
  \bibinfo{author}{\bibfnamefont{D.~L.} \bibnamefont{Graham}},
  \bibinfo{author}{\bibfnamefont{P.~P.} \bibnamefont{Freitas}},
  \bibinfo{author}{\bibfnamefont{L.~A.} \bibnamefont{Clarke}},
  \bibnamefont{and}
  \bibinfo{author}{\bibfnamefont{M.~D.} \bibnamefont{Amaral}},
  \bibinfo{journal}{IEE\ Proc.:\ Circuits\ Devices\ Syst.} \textbf{\bibinfo{volume}{152}},
  \bibinfo{pages}{393} (\bibinfo{year}{2005}).

\bibitem[{\citenamefont{Lee et~al.}(2008)\citenamefont{Lee, Lien, Lee, Lei}}]{Lee2008}
\bibinfo{author}{\bibfnamefont{W.~-C.}~\bibnamefont{Lee}},
  \bibinfo{author}{\bibfnamefont{K.~-Y.}~\bibnamefont{Lien}},
  \bibinfo{author}{\bibfnamefont{G.~-B.} \bibnamefont{Lee}},
  \bibnamefont{and}
  \bibinfo{author}{\bibfnamefont{H.~-Y.} \bibnamefont{Lei}},
  \bibinfo{journal}{Diag.\ Microbiol.\ Infect.\ Dis.} \textbf{\bibinfo{volume}{60}},
  \bibinfo{pages}{51} (\bibinfo{year}{2008}).

\bibitem[{\citenamefont{Pankhurst et~al.}(2003)\citenamefont{Pankhurst, Connolly, Jones, Dobson}}]{Pankhurst2003}
\bibinfo{author}{\bibfnamefont{Q.~A.}~\bibnamefont{Pankhurst}},
  \bibinfo{author}{\bibfnamefont{J.} \bibnamefont{Connolly}},
  \bibinfo{author}{\bibfnamefont{S.~K.}~\bibnamefont{Jones}},
  \bibnamefont{and}
  \bibinfo{author}{\bibfnamefont{J.} \bibnamefont{Dobson}},
  \bibinfo{journal}{J.\ Phys. D:\ Appl.\ Phys.} \textbf{\bibinfo{volume}{36}},
  \bibinfo{pages}{R167} (\bibinfo{year}{2003}).

\bibitem[{\citenamefont{Megens et~al.}(2007)\citenamefont{Megens, de Theije, de Boer, van Gaal}}]{Megens2007}
\bibinfo{author}{\bibfnamefont{M.}~\bibnamefont{Megens}},
  \bibinfo{author}{\bibfnamefont{F.}~\bibnamefont{de Theije}},
  \bibinfo{author}{\bibfnamefont{B.}~\bibnamefont{de Boer}},
  \bibnamefont{and}
  \bibinfo{author}{\bibfnamefont{F.} \bibnamefont{van Gaal}},
  \bibinfo{journal}{J.\ Appl.\ Phys.} \textbf{\bibinfo{volume}{102}},
  \bibinfo{pages}{014507} (\bibinfo{year}{2007}).

\bibitem[{\citenamefont{Ejsing et~al.}(2004)\citenamefont{Ejsing, Hansen, Menon, Ferreira, Graham, Freitas}}]{Ejsing2004}
\bibinfo{author}{\bibfnamefont{L.}~\bibnamefont{Ejsing}},
  \bibinfo{author}{\bibfnamefont{M.~F.}~\bibnamefont{Hansen}},
  \bibinfo{author}{\bibfnamefont{A.~K.} \bibnamefont{Menon}},
  \bibinfo{author}{\bibfnamefont{H.~A.} \bibnamefont{Ferreira}},
  \bibinfo{author}{\bibfnamefont{D.~L.}~\bibnamefont{Graham}},
  \bibnamefont{and}
  \bibinfo{author}{\bibfnamefont{P.~P.} \bibnamefont{Freitas}},
  \bibinfo{journal}{Appl.\ Phys.\ Lett.} \textbf{\bibinfo{volume}{84}},
  \bibinfo{pages}{4729} (\bibinfo{year}{2004}).

\bibitem[{\citenamefont{Li et~al.}(2003)\citenamefont{Li, Joshi, White, Wang, Kemp, Webb, Davis, Sun}}]{Li2003}
\bibinfo{author}{\bibfnamefont{G.}~\bibnamefont{Li}},
  \bibinfo{author}{\bibfnamefont{V.}~\bibnamefont{Joshi}},
  \bibinfo{author}{\bibfnamefont{R.~L.} \bibnamefont{White}},
  \bibinfo{author}{\bibfnamefont{S.} \bibnamefont{Wang}},
  \bibinfo{author}{\bibfnamefont{J.~T.}~\bibnamefont{Kemp}},
  \bibinfo{author}{\bibfnamefont{C.}~\bibnamefont{Webb}},
  \bibinfo{author}{\bibfnamefont{R.~W.}~\bibnamefont{Davis}},
  \bibnamefont{and}
  \bibinfo{author}{\bibfnamefont{S.} \bibnamefont{Sun}},
  \bibinfo{journal}{J.\ Appl.\ Phys.} \textbf{\bibinfo{volume}{93}},
  \bibinfo{pages}{7557} (\bibinfo{year}{2003}).

\bibitem[{\citenamefont{Li et~al.}(2006)\citenamefont{Li, Sun, Wilson, White, Pourmand, Wang}}]{Li2006}
\bibinfo{author}{\bibfnamefont{G.}~\bibnamefont{Li}},
  \bibinfo{author}{\bibfnamefont{S.} \bibnamefont{Sun}},
  \bibinfo{author}{\bibfnamefont{R.~J.} \bibnamefont{Wilson}},
  \bibinfo{author}{\bibfnamefont{R.~L.} \bibnamefont{White}},
  \bibinfo{author}{\bibfnamefont{N.} \bibnamefont{Pourmand}},
  \bibnamefont{and}
  \bibinfo{author}{\bibfnamefont{S.} \bibnamefont{Wang}},
  \bibinfo{journal}{Sens.\ Actuators\ A} \textbf{\bibinfo{volume}{126}},
  \bibinfo{pages}{98} (\bibinfo{year}{2006}).

\bibitem[{\citenamefont{Besse et~al.}(2002)\citenamefont{Besse, Boero, Demierre, Pot, Popovi\'{c}}}]{Besse2002}
\bibinfo{author}{\bibfnamefont{P.~A.}~\bibnamefont{Besse}},
  \bibinfo{author}{\bibfnamefont{G.}~\bibnamefont{Boero}},
  \bibinfo{author}{\bibfnamefont{M.}~\bibnamefont{Demierre}},
  \bibinfo{author}{\bibfnamefont{V.} \bibnamefont{Pot}},
  \bibnamefont{and}
  \bibinfo{author}{\bibfnamefont{R.} \bibnamefont{Popovi\'{c}}},
  \bibinfo{journal}{Appl.\ Phys.\ Lett.} \textbf{\bibinfo{volume}{80}},
  \bibinfo{pages}{4199} (\bibinfo{year}{2002}).

\bibitem[{\citenamefont{Mihajlovi\'{c} et~al.}(2005)\citenamefont{Mihajlovi\'{c}, Xiong, von Moln\'{a}r, Ohtani, Ohno, Field, Sullivan}}]{Mihajlovic2005}
\bibinfo{author}{\bibfnamefont{G.}~\bibnamefont{Mihajlovi\'{c}}},
  \bibinfo{author}{\bibfnamefont{P.}~\bibnamefont{Xiong}},
  \bibinfo{author}{\bibfnamefont{S.} \bibnamefont{Moln\'{a}r}},
  \bibinfo{author}{\bibfnamefont{K.} \bibnamefont{Ohtani}},
  \bibinfo{author}{\bibfnamefont{H.}~\bibnamefont{Ohno}},
  \bibinfo{author}{\bibfnamefont{M.}~\bibnamefont{Field}},\bibnamefont{and}
  \bibinfo{author}{\bibfnamefont{G.~J.} \bibnamefont{Sullivan}},
  \bibinfo{journal}{Appl.\ Phys.\ Lett.} \textbf{\bibinfo{volume}{87}},
  \bibinfo{pages}{112502} (\bibinfo{year}{2005}).

\bibitem[{\citenamefont{Florescu et~al.}(2008)\citenamefont{Florescu, Mattmann, Boser}}]{Florescu2008}
\bibinfo{author}{\bibfnamefont{O.}~\bibnamefont{Florescu}},
  \bibinfo{author}{\bibfnamefont{M.}~\bibnamefont{Mattmann}},
  \bibnamefont{and}
  \bibinfo{author}{\bibfnamefont{B.} \bibnamefont{Boser}},
  \bibinfo{journal}{J.\ Appl.\ Phys.} \textbf{\bibinfo{volume}{103}},
  \bibinfo{pages}{046601} (\bibinfo{year}{2008}).

\bibitem[{\citenamefont{Mihajlovi\'{c} et~al.}(2007)\citenamefont{Mihajlovi\'{c}, Aledealat, Xiong, von Moln\'{a}r, Field, Sullivan}}]{MihajlovicAPL2007}
\bibinfo{author}{\bibfnamefont{G.}~\bibnamefont{Mihajlovi\'{c}}},
  \bibinfo{author}{\bibfnamefont{K.}~\bibnamefont{Aledealat}},
  \bibinfo{author}{\bibfnamefont{P.}~\bibnamefont{Xiong}},
  \bibinfo{author}{\bibfnamefont{S.} \bibnamefont{Moln\'{a}r}},
  \bibinfo{author}{\bibfnamefont{M.}~\bibnamefont{Field}},\bibnamefont{and}
  \bibinfo{author}{\bibfnamefont{G.~J.} \bibnamefont{Sullivan}},
  \bibinfo{journal}{Appl.\ Phys.\ Lett.} \textbf{\bibinfo{volume}{91}},
  \bibinfo{pages}{172518} (\bibinfo{year}{2007}).

\bibitem[{\citenamefont{Choi et~al.}(2002)\citenamefont{Choi, Oh, Thomas, Heineman, Halsall, Nevin, Helmicki, Henderson, Ahn}}]{Choi2002}
\bibinfo{author}{\bibfnamefont{J.~-W.}~\bibnamefont{Choi}},
  \bibinfo{author}{\bibfnamefont{K.~W.}~\bibnamefont{Oh}},
  \bibinfo{author}{\bibfnamefont{J.~H.} \bibnamefont{Thomas}},
  \bibinfo{author}{\bibfnamefont{W.~R.} \bibnamefont{Heineman}},
  \bibinfo{author}{\bibfnamefont{H.~B.} \bibnamefont{Halsall}},
  \bibinfo{author}{\bibfnamefont{J.~H.} \bibnamefont{Nevin}},
  \bibinfo{author}{\bibfnamefont{A.~J.} \bibnamefont{Helmicki}},
  \bibinfo{author}{\bibfnamefont{H.~T.} \bibnamefont{Henderson}},
  \bibnamefont{and}
  \bibinfo{author}{\bibfnamefont{C.~H.} \bibnamefont{Ahn}},
  \bibinfo{journal}{Lab\ Chip} \textbf{\bibinfo{volume}{2}},
  \bibinfo{pages}{27} (\bibinfo{year}{2002}).

\bibitem[{\citenamefont{Janssen et~al.}(2008)\citenamefont{Janssen, van IJzendoorn, Prins}}]{Janssen2008}
\bibinfo{author}{\bibfnamefont{X.~J.~A.}~\bibnamefont{Janssen}},
  \bibinfo{author}{\bibfnamefont{L.~J.}~\bibnamefont{van IJzendoorn}},
  \bibnamefont{and}
  \bibinfo{author}{\bibfnamefont{M.~W.~J.} \bibnamefont{Prins}},
  \bibinfo{journal}{Biosens.\ Bioelectron.} \textbf{\bibinfo{volume}{23}},
  \bibinfo{pages}{833} (\bibinfo{year}{2008}).

\bibitem[{\citenamefont{Spero et~al.}(2008)\citenamefont{Richard Chasen Spero, Leandra Vicci,2 Jeremy Cribb,3 David Bober,4 Vinay Swaminathan,5 E. Timothy O'Brien,1 Stephen L. Rogers,6 and R. Superfine1}}]{Spero2008}
\bibinfo{author}{\bibfnamefont{R.~C.}~\bibnamefont{Spero}},
  \bibinfo{author}{\bibfnamefont{L.}~\bibnamefont{Vicci}},
  \bibinfo{author}{\bibfnamefont{J.} \bibnamefont{Cribb}},
  \bibinfo{author}{\bibfnamefont{D.} \bibnamefont{Bober}},
  \bibinfo{author}{\bibfnamefont{V.}~\bibnamefont{Swaminathan}},
  \bibinfo{author}{\bibfnamefont{E.~T.}~\bibnamefont{O'Brien}},
  \bibinfo{author}{\bibfnamefont{S.~L.}~\bibnamefont{Rogers}},
  \bibnamefont{and}
  \bibinfo{author}{\bibfnamefont{R.} \bibnamefont{Superfine}},
  \bibinfo{journal}{Rev.\ Sci.\ Instrum.} \textbf{\bibinfo{volume}{79}},
  \bibinfo{pages}{083707} (\bibinfo{year}{2008}).

\bibitem[{\citenamefont{Weddemann et~al.}(2009)\citenamefont{Weddemann, Wittbracht, Auge, H\"{u}tten}}]{Weddemann2009}
\bibinfo{author}{\bibfnamefont{A.}~\bibnamefont{Weddemann}},
  \bibinfo{author}{\bibfnamefont{F.}~\bibnamefont{Wittbracht}},
  \bibinfo{author}{\bibfnamefont{A.}~\bibnamefont{Auge}},
  \bibnamefont{and}
  \bibinfo{author}{\bibfnamefont{A.} \bibnamefont{H\"{u}tten}},
  \bibinfo{journal}{Appl.\ Phys.\ Lett.} \textbf{\bibinfo{volume}{94}},
  \bibinfo{pages}{173501} (\bibinfo{year}{2009}).

\bibitem[{\citenamefont{Weibel et~al.}(2006)\citenamefont{Weibel, Whitesides}}]{Weibel2006}
\bibinfo{author}{\bibfnamefont{D.~B.}~\bibnamefont{Weibel}}
  \bibnamefont{and}
  \bibinfo{author}{\bibfnamefont{G.~M.} \bibnamefont{Whitesides}},
  \bibinfo{journal}{Curr.\ Opin.\ Chem.\ Biol.} \textbf{\bibinfo{volume}{10}},
  \bibinfo{pages}{584} (\bibinfo{year}{2006}).

\bibitem[{\citenamefont{Loureiro et~al.}(2009)\citenamefont{Loureiro, Ferreira, Cardoso, Freitas, Germano, Fermon, Arrias, Pannetier-Lecoeur, Rivadulla, Rivas}}]{Loureiro2009}
\bibinfo{author}{\bibfnamefont{J.}~\bibnamefont{Loureiro}},
  \bibinfo{author}{\bibfnamefont{R.}~\bibnamefont{Ferreira}},
  \bibinfo{author}{\bibfnamefont{S.} \bibnamefont{Cardoso}},
  \bibinfo{author}{\bibfnamefont{P.~P.} \bibnamefont{Freitas}},
  \bibinfo{author}{\bibfnamefont{J.}~\bibnamefont{Germano}},
  \bibinfo{author}{\bibfnamefont{C.}~\bibnamefont{Fermon}},
  \bibinfo{author}{\bibfnamefont{G.}~\bibnamefont{Arrias}},
  \bibinfo{author}{\bibfnamefont{M.}~\bibnamefont{Pannetier-Lecoeur}},
  \bibinfo{author}{\bibfnamefont{F.}~\bibnamefont{Rivadulla}},
  \bibnamefont{and}
  \bibinfo{author}{\bibfnamefont{J.} \bibnamefont{Rivas}},
  \bibinfo{journal}{Appl.\ Phys.\ Lett.} \textbf{\bibinfo{volume}{95}},
  \bibinfo{pages}{034104} (\bibinfo{year}{2009}).

\bibitem[{\citenamefont{Liu et~al.}(2009)\citenamefont{Liu, Stakenborg, Peeters, Lagae}}]{Liu2009}
\bibinfo{author}{\bibfnamefont{C.}~\bibnamefont{Liu}},
  \bibinfo{author}{\bibfnamefont{T.}~\bibnamefont{Stakenborg}},
  \bibinfo{author}{\bibfnamefont{S.} \bibnamefont{Peeters}},
  \bibnamefont{and}
  \bibinfo{author}{\bibfnamefont{L.} \bibnamefont{Lagae}},
  \bibinfo{journal}{J.\ Appl.\ Phys.} \textbf{\bibinfo{volume}{105}},
  \bibinfo{pages}{102014} (\bibinfo{year}{2009}).

\bibitem[{\citenamefont{Shen et~al.}(2005)\citenamefont{Shen, Liu, Mazumdar, Xiao}}]{Shen2005}
\bibinfo{author}{\bibfnamefont{W.}~\bibnamefont{Shen}},
  \bibinfo{author}{\bibfnamefont{X.}~\bibnamefont{Liu}},
  \bibinfo{author}{\bibfnamefont{D.} \bibnamefont{Mazumdar}},
  \bibnamefont{and}
  \bibinfo{author}{\bibfnamefont{G.} \bibnamefont{Xiao}},
  \bibinfo{journal}{Appl.\ Phys.\ Lett.} \textbf{\bibinfo{volume}{86}},
  \bibinfo{pages}{253901} (\bibinfo{year}{2005}).

\bibitem[{\citenamefont{Thilwind et~al.}(2008)\citenamefont{Thilwind, Megens, van Zon, Coehoorn, Prins}}]{Thilwind2008}
\bibinfo{author}{\bibfnamefont{R.~E.}~\bibnamefont{Thilwind}},
  \bibinfo{author}{\bibfnamefont{M.}~\bibnamefont{Megens}},
  \bibinfo{author}{\bibfnamefont{J.~B.~A.~D.} \bibnamefont{van Zon}},
  \bibinfo{author}{\bibfnamefont{R.}~\bibnamefont{Coehoorn}},
  \bibnamefont{and}
  \bibinfo{author}{\bibfnamefont{M.~W.~J.} \bibnamefont{Prins}},
  \bibinfo{journal}{J.\ Magn.\ Magn.\ Mater.} \textbf{\bibinfo{volume}{320}},
  \bibinfo{pages}{486} (\bibinfo{year}{2008}).

\bibitem[{\citenamefont{Roh et~al.}(2009)\citenamefont{Roh, Son, Lee, Lee, Jung, Lee}}]{Roh2009}
\bibinfo{author}{\bibfnamefont{J.~W.}~\bibnamefont{Roh}},
  \bibinfo{author}{\bibfnamefont{O.}~\bibnamefont{Son}},
  \bibinfo{author}{\bibfnamefont{Y.~T.} \bibnamefont{Lee}},
  \bibinfo{author}{\bibfnamefont{K.}~\bibnamefont{Lee}},
  \bibinfo{author}{\bibfnamefont{H.}~\bibnamefont{Jung}},
  \bibnamefont{and}
  \bibinfo{author}{\bibfnamefont{W.} \bibnamefont{Lee}},
  \bibinfo{journal}{Phys.\ Stat.\ Sol.\ A} \textbf{\bibinfo{volume}{206}},
  \bibinfo{pages}{1636} (\bibinfo{year}{2009}).

\bibitem[{\citenamefont{Osterberg et~al.}(2009)\citenamefont{Osterberg, Dalslet, Damsgaard, Freitas, Freitas, Hansen}}]{Osterberg2009}
\bibinfo{author}{\bibfnamefont{J.~W.}~\bibnamefont{{\O}sterberg}},
  \bibinfo{author}{\bibfnamefont{O.}~\bibnamefont{Dalslet}},
  \bibinfo{author}{\bibfnamefont{Y.~T.} \bibnamefont{Damsgaard}},
  \bibinfo{author}{\bibfnamefont{K.}~\bibnamefont{Freitas}},
  \bibinfo{author}{\bibfnamefont{H.}~\bibnamefont{Freitas}},
  \bibnamefont{and}
  \bibinfo{author}{\bibfnamefont{W.} \bibnamefont{Hansen}},
  \bibinfo{journal}{IEEE\ Sens.\ J.} \textbf{\bibinfo{volume}{9}},
  \bibinfo{pages}{682} (\bibinfo{year}{2009}).

\bibitem[{\citenamefont{Manandhar et~al.}(2009)\citenamefont{Manandhar, Chen, Aledealat, Mihajlovi\'{c}, Yun, Field, Sullivan, Strouse, Chase, von Moln\'{a}r, Xiong}}]{Manandhar2009}
\bibinfo{author}{\bibfnamefont{P.}~\bibnamefont{Manandhar}},
  \bibinfo{author}{\bibfnamefont{K.~-S.}~\bibnamefont{Chen}},
  \bibinfo{author}{\bibfnamefont{K.}~\bibnamefont{Aledealat}},
  \bibinfo{author}{\bibfnamefont{G.} \bibnamefont{Mihajlovi\'{c}}},
  \bibinfo{author}{\bibfnamefont{C.~S.}~\bibnamefont{Yun}},
  \bibinfo{author}{\bibfnamefont{M.}~\bibnamefont{Field}},
  \bibinfo{author}{\bibfnamefont{G.~J.}~\bibnamefont{Sullivan}},
  \bibinfo{author}{\bibfnamefont{G.~F.}~\bibnamefont{Strouse}},
  \bibinfo{author}{\bibfnamefont{P.~B.} \bibnamefont{Chase}},
  \bibinfo{author}{\bibfnamefont{S.}~\bibnamefont{von Moln\'{a}r}},
  \bibnamefont{and}
  \bibinfo{author}{\bibfnamefont{P.} \bibnamefont{Xiong}},
  \bibinfo{journal}{Nanotechnology} \textbf{\bibinfo{volume}{20}},
  \bibinfo{pages}{355501} (\bibinfo{year}{2009}).

\bibitem[{\citenamefont{Mihajlovi\'{c} et~al.}(2007)\citenamefont{Mihajlovi\'{c}, Xiong, von Moln\'{a}r, Field, Sullivan}}]{MihajlovicJAP2007}
\bibinfo{author}{\bibfnamefont{G.}~\bibnamefont{Mihajlovi\'{c}}},
  \bibinfo{author}{\bibfnamefont{P.}~\bibnamefont{Xiong}},
  \bibinfo{author}{\bibfnamefont{S.} \bibnamefont{Moln\'{a}r}},
  \bibinfo{author}{\bibfnamefont{M.}~\bibnamefont{Field}},
  \bibnamefont{and}
  \bibinfo{author}{\bibfnamefont{G.~J.} \bibnamefont{Sullivan}},
  \bibinfo{journal}{J.\ Appl.\ Phys.} \textbf{\bibinfo{volume}{102}},
  \bibinfo{pages}{034506} (\bibinfo{year}{2007}).

\end{thebibliography}

\end{document}